\documentclass[epj-spec]{svjour}
\usepackage{graphics}
\usepackage[]{amsmath}
\usepackage{url}
\begin{document}
%
\title{Multiconfiguration Dirac-Hartree-Fock calculations of
 transition rates 
 and lifetimes of the eight lowest excited levels of radium}
%

\author{
 Jacek Biero\'n\inst{1}
\and
 Paul Indelicato\inst{2}
\and 
 Per J{\"o}nsson\inst{3}}
\institute{Instytut Fizyki imienia~Mariana~Smoluchowskiego \\
 Uniwersytet Jagiello{\'n}ski \\
 Reymonta~4, 30-059~Krak{\'o}w, Poland \\
 \email{Bieron@uj.edu.pl}
\and 
Laboratoire Kastler Brossel,
\'Ecole Normale Sup\' erieure \\
 CNRS; Universit\' e P.~et M.~Curie - Paris 6 \\
Case 74; 4, place Jussieu, 75252 Paris CEDEX 05, France \\
\email{paul.indelicato@spectro.jussieu.fr}
\and 
Nature, Environment, Society \\
Malm\"o University\\
S-205~06 Malm{\"o}, Sweden\\ 
\email{per.jonsson@ts.mah.se}
}



%
\abstract{
The multiconfiguration Dirac-Hartree-Fock (MCDHF) model has been employed
to calculate the transition rates between the nine lowest levels
of radium.
The dominant rates were then used to evaluate the 
radiative lifetimes. The decay of the metastable $ 7s7p \, ^3 \! P _0 $ state
through 2-photon E1M1 and hyperfine induced channels is also studied.
} 
\maketitle
%
\section{Introduction}
\label{Introduction}

%
Recent advances in trapping and spectroscopy of free, neutral
atoms make it possible to extend the
search for time reversal violation effects 
into the domain of radioactive elements~\cite{Behr2005}.
In the last decade several heavy atoms
were considered
as candidates for experimental searches~\cite{Ginges2003}.
%
There are at least two ongoing atomic trap experiments
(in Kernfysisch Versneller Instituut in
Netherlands~\cite{Jungmann2002,edmKVI}
 and in Argonne National Lab in 
 the~U.S.~\cite{Scielzo2006,edmArgonne})
whose  aim is to detect the electric dipole moment of radium.
The advantage of radium lies in
octupole deformations of nuclei in several isotopes~\cite{Engel2005},
simple electronic structure
(ground state configuration
 [Kr]$ 4d^{10} 4f^{14} 5s^2 5p^6 5d^{10} 6s^2 6p^6 7s^2$
 yields the closed-shell singlet state $^1 \! S _0$)
as well as in
coincidental proximity of two atomic states of opposite parity,
 $ 7s7p \, ^3 \! P _1 $ and $ 7s6d \, ^3 \! D _2 $, which are separated
by a very small energy interval 5~cm$^{-1}$.

The data on atomic spectrum of radium compiled in the tables
of Moore~\cite{Moore1971} came from the experimental investigations
of Rasmussen~\cite{Rasmussen1934}, with subsequent revisions
by Russell~\cite{Russell1934}; both go back to the 1930s.
These data cover only 69 classified lines.
%
The isotope shifts and hyperfine structures of radium
were measured by the group of
Wendt~\cite{Ahmad1983,Wendt1987,Arnold1987,Ahmad1988,Neu1989,Neugart1990}
in the 1980s.
 They have studied both atomic Ra~I and ionic Ra~II spectra,
 and obtained the isotope shifts,
 magnetic dipole hyperfine constants $A$ and
 electric quadrupole constants $B$
 of the
 $7s7p \, ^1 \! P _1$,
 $7s7p \, ^3 \! P _1$,
 $7s7p \, ^3 \! P _2$, and
 $7s7d \, ^3 \! D _3$ levels of neutral radium,
%
 as well as the magnetic dipole hyperfine constants $A$
 of the
 $7s \, ^2 \! S _{1/2}$,
 $7p \, ^2 \! P _{1/2}$, and
 $7p \, ^2 \! P _{3/2}$ levels of Ra$^+$~ion,
 together with the electric quadrupole constant $B$
 of the
 $7p \, ^2 \! P _{3/2}$ state of Ra$^+$.
%
Hyperfine structures of singly-ionised radium have been also the subject
of several theoretical
papers~\cite{Dzuba1985,Heully1985,Panigrahy1991,Andriessen1992,%
YuanCa1995,YuanRa1995}.
The excitation energies of several states of
neutral and singly-ionised radium (and barium) were
later calculated in the framework the relativistic coupled-cluster theory
by the group of Kaldor~\cite{Eliav1996,Landau2000}.
More recently there have been three papers from our
group 
in which we calculated
the lifetimes of the 
$7s7p \, ^1 \! P _1$ and
$7s6d \, ^3 \! D _2$ states~\cite{BieronRa3d2004};
the hyperfine structure constants
of all levels belonging to the two lowest excited-state
configurations~\cite{BieronRahfs2005};
 and the 
 electric field gradients 
 generated by the electronic cloud
 in the
 $7s7p \, ^1 \! P _1$,
 $7s7p \, ^3 \! P _1$, and
 $7s7p \, ^3 \! P _2$
 states, which in turn
 (combined with the measured values of the
 electric quadrupole constants $B$)
 yielded the nuclear electric quadrupole moment
of radium-223 isotope~\cite{BieronRaQ2005}.
The excitation energies and lifetimes 
of several states of radium (and barium) were
calculated recently
in the framework of the combined configuration-interaction
and many-body perturbation theory
by Dzuba and Ginges~\cite{DzubaGinges2006}.
 
In the present paper we calculated the 
transition probabilities between the states
arising from the three lowest configurations of radium:
%
$ 7s^2~ ^1 \! S _0$,
$ 7s7p~ ^3 \! P _{0,1,2}$,
$ 7s7p~ ^1 \! P _1$,
$ 7s6d~ ^3 \! D _{1,2,3}$, and
$ 7s6d~ ^1 \! D _2$,
as well as the lifetimes of these states.
The purpose of the present paper is fourfold.
Firstly, we intended to extend the transition rate
calculations on all levels arising from the three lowest
configurations (i.e. $7s^2$, $7s7p$, and $7s6d$).
We included
(1) transitions which contribute appreciably to lifetimes;
(2) transitions involving
$ 7s6d~ ^3 \! D _{2}$ and
$ 7s7p~ ^3 \! P _{1}$
 levels because they are of interest
    in EDM experiments (the mixing
    induced by T-odd interactions is strongest between these two levels);
(3) transitions which may be important for trapping;
(4) transitions which are stronger than $1/s$ (a somewhat arbitrary threshold).
%
Secondly,
new comparison of our results became available
when the two most recent papers~\cite{DzubaGinges2006,Scielzo2006}
 appeared in print.
%
These in turn permitted further tests of
 the newly developed~\cite{grasp2K}
parallel version of the
GRASP package~\cite{grasp92}
and calibration the theoretical model for the calculations
of the spectroscopic properties of radium.
Finally, we present a summary 
of the available theoretical data on the 
spectroscopic properties
of eight lowest excited levels of radium,
with the hope that they may be
of help for the experimental groups, that are currently
in the process of setting up the atomic traps for the search of
permanent electric dipole moments.
The Argonne group has already measured~\cite{Scielzo2006}
the frequency and
the rate of the
$ ^3 \! P _1 - \, ^1 \! S _0 $
transition in 
the $^{225}_{\phantom{1}88}$Ra isotope,
and determined the  lifetime of the $ ^3 \! P _1 $ level.
The lifetime of this level have been previously
calculated by Hafner and Schwarz~\cite{HafnerSchwarz1978}
and by Bruneau~\cite{Bruneau1984}.

%
\section{Theory}
\label{Theory}

%
%
The modified version~\cite{grasp2K}
of the GRASP implementation~\cite{grasp92}
of the multiconfiguration Dirac-Hartree-Fock method~\cite{Grant1994}
was used in the present paper.
The starting point is the Dirac-Coulomb Hamiltonian
\begin{equation}
\label{Dirac-Coulomb-Hamiltonian}
H_{DC} = \sum_{i} c {\boldsymbol{ \alpha }}_i \cdot
                    {\boldsymbol{ p }}_i
         + (\beta_i -1)c^2 + V^N_i
         + \sum_{i>j} 1/r_{ij},
\end{equation}
where $V^N$ is the monopole part of the electron-nucleus Coulomb interaction.
The wavefunction for a particular atomic state ($\Psi$)
is obtained as the self-consistent solution of
the Dirac-Fock equation~\cite{grasp92}
in a basis of symmetry adapted
configuration state functions ($\Phi$)
\begin{equation}
\label{ASF}
\Psi({\mathit \Gamma} PJM) = \sum_{i}^{NCF} c_{i} \Phi(\gamma_{i}PJM).
\end{equation}
The basis $NCF$ was systematically
enlarged~\cite{Bieron1999BeF,Bieron2001Bi} to yield increasingly accurate
approximations to the exact wavefunction.
%
%
All calculations were done with
the nucleus modeled as a variable-density sphere,
where a two-parameter Fermi function~\cite{grasp89}
was employed to approximate the charge distribution.
The Breit and QED corrections were estimated
with the step-wise procedure described in~\cite{Bieron1999BeF}.
They were applied only to two transition rates,
as explained in section~\ref{3D2:1P1} below.

\section{Method}
\label{Method}

%
%
The wavefunctions were obtained with the active space method
in which configuration state functions of a particular parity and
symmetry are generated by substitutions from
a reference configuration to an active set of orbitals.
The active set and the
multiconfiguration expansions are increased systematically.
The whole process is governed by
convergence of the expectation values.
%
%
The calculations were divided into two stages. Each stage was
further divided into several consecutive steps.
In the first stage, the spectroscopic and virtual spinorbitals
 were generated in relatively small multiconfiguration expansions.
%
The spectroscopic orbitals
required to form a reference wavefunction
were obtained with a minimal configuration expansion,
with full relaxation.
%
%
Then virtual orbitals were generated
in five consecutive steps.
At each step the virtual set has been extended by one layer
of virtual orbitals. A layer is defined as a set of virtual orbitals
with different angular symmetries. In the present paper
five layers of virtual orbitals of each of the
 {\sl s, p, d, f, g, h} symmetries were generated.
%
At each step the configuration expansions were limited
to single and double
substitutions from valence shells to all new orbitals
and to all previously generated virtual layers.
These were augmented by small subsets of dominant single and double
substitutions from core and valence shells, with further
restriction, that at most one electron may be
promoted from core shells (which means, that in the case
of a double substitution the second electron must be promoted
from a valence shell).
All configurations from earlier steps
were retained, with all previously generated orbitals
fixed, and all new orbitals made orthogonal to others of the
same symmetry.
%
%
The initial shapes of radial orbitals were obtained
in Thomas-Fermi potential, and then driven to convergence with
the self-consistency threshold set to $10^{-8}$.
%
All radial orbitals 
were separately optimized for each of the nine atomic
states of interest.
The Optimal Level
form of the variational expression~\cite{grasp89}
was applied in all variational calculations.

%
%
In the second stage, the configuration-interaction calculations
(i.e.,~with no changes to the radial wavefunctions) 
were performed, with multiconfiguration expansions
tailored in such a way, as to capture the dominant electron
correlation contributions to the expectation values.
The valence and core-valence effects
constitute the dominant electron correlation contributions in
the oscillator strength calculations~\cite{FBJbook.p71},
 therefore
all single and double substitutions were allowed from several core
shells and both valence shells
(i.e.,~$7s^2$, $7s7p$, or $7s6d$, depending on the state)
to all virtual shells, with the same restriction
as above, i.e.~that at most one electron may be promoted from core shells.
The virtual set was systematically increased from one to five
layers,
until the convergence of transition rates was obtained.
In a similar manner, several core shells were systematically
opened for electron substitutions --- from the outermost $6p$
to $5s 5p 5d 6s 6p$ shells.
The effects of substitutions from
$4s4p4d4f$ shells were neglected.
We estimated them separately for three states
and discovered that they
change the calculated values of transition rates
by no more than a fraction of a percent.
%
%
The transition rates were calculated with the biorthonormal
technique~\cite{Malmqvist1986,Olsen1995}, which permits the
application of standard Racah algebra, while retaining the
advantage of wavefunctions separately optimized for each state.
Experimental values of transition energies from
 Moore's tables~\cite{Moore1971}
were used in calculations of transition rates.

\section{Results}
\label{Results}

%
\subsection{The metastable $7s7p \, ^3 \! P _0$ state}
\label{3P0}

In principle there are three possible decay channels of the
 $7s7p \, ^3 \! P _0$ state.
It can decay to (the only lower lying) ground state through 
(1) a blackbody radiation induced decay,
(2) a 2-photon E1M1 transition, or through 
(3) a hyperfine induced transition.
The first is beyond the scope of the present paper,
since it depends on the ambient temperature.
Of the other two,
the former
can be estimated through an order-of-magnitude
comparison with the E1M1 
$1s2p \, ^3 \! P _0$~--~$1s^2 \, ^1 \! S _0$ two-photon transition
 in helium-like heavy ions. 
We started from a recent
evaluation~\cite{SavukovJohnson2002}, which gives
$3.14\times 10^9$~s$^{-1}$ for Ra.
In order to obtain a good dependence on 
 transition energy and radial matrix elements, we evaluated
the E1M1 matrix elements
(following Eq.~(2) of Ref.~\cite{SavukovJohnson2002}) for He-like Ra
(using only $1s2p\,^3 \! P_1$ as an intermediate state)
and for neutral Ra (using only the $7s7p \,^3 \! P_1$).
Then we integrated over photon energies with the five-point
Gauss-Legendre formula.
Finally, we evaluated the ratio of these two values to scale
 the He-like E1M1 rate.
This gives an order of magnitude estimate of $9.6\times 10^{-3}$~$s^{-1}$.
%
%
Such lifetime ($\approx 100$~s)  of the $7s7p \, ^3 \! P _0$ state
would be comparable to the lifetimes of nuclei of
several radium isotopes
(it would be significantly shorter only in comparison with
the nuclei of the most stable radium isotopes
spanning the mass range 223-229).
%
%
This would be the case of spin-zero isotopes of radium.

%
For the isotopes of radium with a nonzero value
of nuclear spin,
the hyperfine-induced transition must also be considered.
We estimated the $ 7s7p \, ^3 \! P _0 $~--~$ 7s^2 \, ^1 \! S _0 $
transition rate with a simple
three-state model, in which the wavefunctions
of the hyperfine components of the upper $^3 \! P _0$
state are described by a
symmetry-adapted configuration-state-function expansion
of the form 
\begin{equation}
\label{HFSexpansion}
     \vert 7s7p~ ^3 \! P _0 I F \rangle _{HFS} =
 c_0 \vert 7s7p~ ^3 \! P _0 I F \rangle +
 c_1 \vert 7s7p~ ^1 \! P _1 I F \rangle +
 c_2 \vert 7s7p~ ^3 \! P _1 I F \rangle 
\end{equation}
running over the appropriate hyperfine components $IF$ of the 
$^1 \! P _1$ and
$^3 \! P _1$
states.
The hyperfine-induced
transition rate (in $s^{-1}$) 
may be approximately expressed as
\begin{eqnarray}
\label{A-hyperfine-induced}
A(7s7p~^3 \! P_0^o \to 7s^2~^1 \! S_0) = \nonumber \\
\frac{2.02613 \times 10^{18}}{3 \lambda^3}
\left|  c_1 \langle  7s7p~^1 \! P_1^o \| {\bf Q}^{(1)}_{1} 
                  \| 7s^2~^1 \! S_0 \rangle +
        c_2 \langle  7s7p~^3 \! P_1^o \| {\bf Q}^{(1)}_{1}
                  \| 7s^2~^1 \! S_0  \right|^2   
\end{eqnarray}
where
  $\langle  7s7p~^1 \! P_1^o \| {\bf Q}^{(1)}_{1} \| 7s^2~^1 \! S_0 \rangle$
 and
  $\langle  7s7p~^3 \! P_1^o \| {\bf Q}^{(1)}_{1} \| 7s^2~^1 \! S_0 \rangle$
are reduced matrix elements for the electric dipole operator, and
$\lambda$ is the transition wavelength (in {\AA}).
%
%
%
The coefficients 
$ c_1 $ and
$ c_2 $
are related to the off-diagonal magnetic dipole constants
$A ^{HFS} _{M1}$($ ^1 \! P _1, ^3 \! \! P _0 $)
and
$A ^{HFS} _{M1}$($ ^3 \! P _1, ^3 \! \! P _0 $)
\begin{equation}
\label{c12}
 c_1 = \sqrt{I(I+1)}
       \frac{ A^{HFS}_{M1} ( ^1 \! P _1,  ^3 \! \! P _0 ) }
            {\Delta E(^1 \! P _1 - \,  ^3 \! P _0)} ,
\ \ \ \ \
 c_2 = \sqrt{I(I+1)}
       \frac{ A^{HFS}_{M1} ( ^3 \! P _1,  ^3 \! \! P _0 ) }
            {\Delta E(^3 \! P _1 - \,  ^3 \! P _0)}
\end{equation}
(see~\cite{Johnson1997} or~\cite{PorsevDerevianko2004}
 for full derivation).
The sum in the expansion~(\ref{HFSexpansion})
should run over all excited states, but the sum in
Eq.~\eqref{A-hyperfine-induced} is usually
dominated by those states, for which the 
transition rates to the ground state are large and at the same
time the energy denominators 
in Eq.~\eqref{c12} are small.
In case of the $ 7s7p~ ^3 \! P _0 $ state of radium the
$ 7s7p~ ^3 \! P _1$ and $ 7s7p~ ^1 \! P _1$ states dominate.
The contribution of the reduced matrix element
  $\langle  7s7p~^3 \! P_1^o \| {\bf Q}^{(1)}_{1} \| 7s^2~^1 \! S_0 \rangle$ 
is 3.7~times larger than that of the
  $\langle  7s7p~^1 \! P_1^o \| {\bf Q}^{(1)}_{1} \| 7s^2~^1 \! S_0 \rangle$
matrix element.
The contributions of other states are much smaller,
due to the presence of the energy denominators 
in the coefficients~(\ref{c12}) and to the fact that the corresponding
off-diagonal matrix elements are many orders of magnitude smaller
(actually they are exactly zero, unless non-orthogonality between fully
relaxed wavefunctions and correlation effects are not neglected),
and they can be safely ignored
at present level of the 
overall accuracy.
%
The rate of the hyperfine-induced transition is isotope-dependent,
i.e.~it depends 
on the nuclear spin and on the nuclear magnetic moment.
As in our previous
paper~\cite{BieronRaQ2005},
the $^{223}_{\phantom{1}88}$Ra isotope
was chosen to set the
nuclear parameters
(the transition rate
$A$($ ^3 \! P _0 - ^1 \! \! S _0 $)
may be readily recalculated for other radium
 isotopes, for which nuclear spins and magnetic moments are known).
The nuclear spin of $^{223}_{\phantom{1}88}$Ra is $ I=3/2 $,
and the nuclear magnetic dipole moment
$ \mu = 0.2705(19) \mu_N $
was taken from
the paper of Arnold~{\sl et al}~\cite{Arnold1987}.
The electric dipole transition rates 
     $A$($ ^1 \! P _1 - \, ^1 \! S _0$)
 and $A$($ ^3 \! P _1 - \, ^1 \! S _0$)
from table~\ref{tableABC} (in Babushkin gauge)
were used.
The  off-diagonal magnetic dipole constants
       $A ^{HFS}_{M1}$($ ^1 \! P _1,  ^3 \! \! P _0 $)~=~540~MHz, and
       $A ^{HFS}_{M1}$($ ^3 \! P _1,  ^3 \! \! P _0 $)~=~1172~MHz,
were evaluated with the use of the same wavefunctions,
and in the same approximation, as described in section~(\ref{Method})
above.
Together they yield hyperfine-induced transition rate 
$A$(\mbox{$ ^3 \! P _0 - \, ^1 \! \! S _0 $})~=~0.0210~s$^{-1}$
in case of constructive interference, and
$A$(\mbox{$ ^3 \! P _0 - \, ^1 \! \! S _0 $})~=~0.0070~s$^{-1}$
in case of destructive interference.

%
The above evaluations of the E1M1 and hyperfine-induced rates
were performed independently. As it turned out,
the two contributions are of the same size,
therefore they have to be treated simultaneously.
To this end we employed
the effective Hamiltonian method \cite{Indelicato1989}.
It requires also an evaluation of
the $A ^{HFS} _{M1}$($ ^3 \! P _1, ^1 \! \! P _1 $)
 matrix element (851~MHz).
The method is valid beyond the limits of the perturbation method exposed
above, in particular it does not require that the energy separation
between levels is large compared to level widths.
It yields $A\left({ ^3 \! P _0 - \, ^1 \! \! S _0 }\right) = 0.02935$~$s^{-1}$
(if the two-photon transition is switched off
the effective Hamiltonian method
 yields $A\left({ ^3 \! P _0 - \, ^1 \! \! S _0 }\right) = 0.0197~s^{-1}$).
In the process of this calculation, we found out by comparison with
the results of \emph{mdfgme} code~\cite{iad2005} that
non-orthogonality between spinorbitals
plays very important role
in the evaluation of non-diagonal hyperfine matrix elements
(neglecting it
may even lead to a sign change).
We employed a new code, developed by one of us (P.J.) to evaluate
correlated off-diagonal
hyperfine matrix elements from GRASP wavefunctions,
taking into account the effect of non-orthogonality between spinorbitals.

%
The accuracy is limited by the electric dipole matrix elements
  $\langle  7s7p~^3 \! P_1^o \| {\bf Q}^{(1)}_{1} \| 7s^2~^1 \! S_0 \rangle$
and
  $\langle  7s7p~^1 \! P_1^o \| {\bf Q}^{(1)}_{1} \| 7s^2~^1 \! S_0 \rangle$
as well as by the 
off-diagonal magnetic dipole constants
$A ^{HFS} _{M1}$($ ^3 \! P _1, ^1 \! \! P _1 $),
       $ A ^{HFS} _{M1}$($ ^3 \! P _1,  ^3 \! \! P _0 $) and
       $ A ^{HFS} _{M1}$($ ^1 \! P _1,  ^3 \! \! P _0 $).
%
In case of the
matrix element
  $\langle  7s7p~^3 \! P_1^o \| {\bf Q}^{(1)}_{1} \| 7s^2~^1 \! S_0 \rangle$ 
we may assume
the 5\%~relative accuracy for the
$ ^3 \! P _1 - \,  ^1 \! S _0 $ transition rate
from experiment~\cite{Scielzo2006}.
The accuracy of the $A$($ ^1 \! P _1 - \, ^1 \! \! S _0$) rate is 
more difficult to estimate (see ref.~\cite{BieronRa3d2004}).
This is the strongest, 'allowed' transition in the radium spectrum,
 but (as mentioned above) the contribution of its
matrix element
  $\langle  7s7p~^1 \! P_1^o \| {\bf Q}^{(1)}_{1} \| 7s^2~^1 \! S_0 \rangle$ 
to the total value of the calculated rate of the
hyperfine-induced transition 
$  ^3 \! P _0 - \, ^1 \! \! S _0 $
is 3.7~times smaller than
that of the
  $\langle  7s7p~^3 \! P_1^o \| {\bf Q}^{(1)}_{1} \| 7s^2~^1 \! S_0 \rangle$
matrix element,
so even a relatively large error bar would be quenched. 
We may very conservatively take the entire difference
between the Babushkin and Coulomb gauge final values 
from table~1 in reference~\cite{BieronRa3d2004}
as the error limit, obtaining a relative accuracy 25\%
for the $A$($ ^1 P _1 - \, ^1 \! S _0$) transition probability.
The accuracy of the diagonal magnetic dipole constants was estimated
to be 6\% (see \cite{BieronRahfs2005}).
Since an off-diagonal constant depends on both states rather than on one,
we, again quite conservatively, doubled the 'diagonal' error limit estimate
and assumed 12\% as the contributions of 
the off-diagonal hyperfine constants
to the error bar.
%
Eventually, the above procedure
yields the lifetime of the metastable 
$ ^3 \! P _0 $ state
$\tau$~=~34(15)$s$ of $^{223}_{\phantom{1}88}$Ra isotope.
The lifetime is based on both the 2-photon and hyperfine-induced
channels. The error bar does not include the 2-photon contribution.

\subsection{The $ 7s7p~ ^3 \! P _1 $ state}
\label{3P1}

\begin{figure}
%
\resizebox{0.8\columnwidth}{!}{%
  \includegraphics{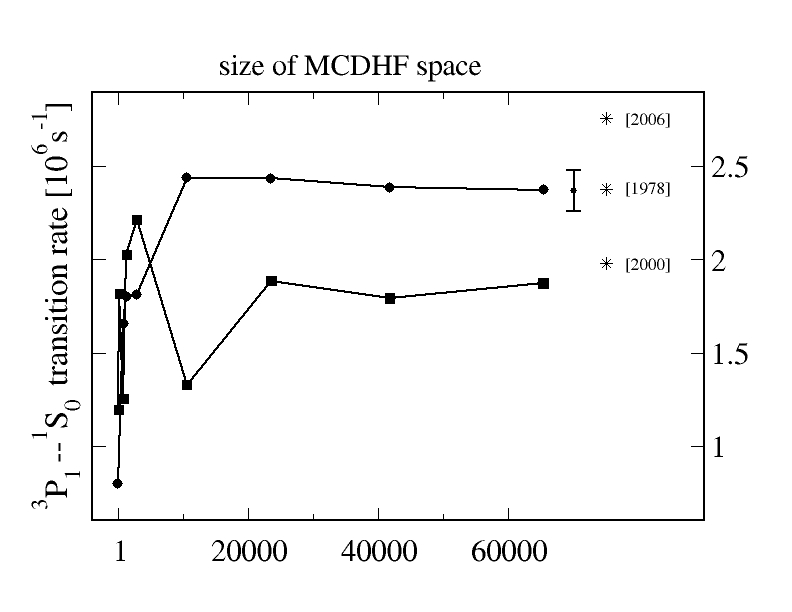} }
\caption{Transition probability $A$($ ^3 \! P _1 - ^1 \! S_0 $)
 in Babushkin (upper curve with circles)
 and Coulomb (lower curve with squares) gauges,
as a function of the multiconfiguration expansion
(in core-valence approximation 
--- see text for details).
The lone dot with error bars 
 (to the right from the end of the Babushkin curve)
represents the experimental result
from reference~\cite{Scielzo2006}.
The three stars at the far right represent the theoretical
data obtained by other authors
and are denoted by publication year in brackets:
[1978] --- reference~\cite{HafnerSchwarz1978};
[2000] --- reference~\cite{Dzuba2000}; 
[2006] --- reference~\cite{DzubaGinges2006}. } 
\label{fig3P1}       
\end{figure} 

Figure~\ref{fig3P1} presents 
the transition probability
$A$($ ^3 \! P _1 - ^1 \! \! S _0 $)
as a function of the size of the multiconfiguration expansion.
The transition rates were calculated in Babushkin and Coulomb gauges,
which in the non-relativistic formulation correspond to length and
velocity form of the transition integral, respectively.
Both curves were obtained
in the 'core-valence' approximation described above.
The resulting Babushkin and Coulomb values are compared
with the experimental result obtained by the
 Argonne group~\cite{Scielzo2006}
and with three available theoretical values
(the value $A$=$4 \cdot 10^{6} s^{-1} $
 obtained by Bruneau~\cite{Bruneau1984}
did not fit within the vertical scale of figure~\ref{fig3P1}).
It is clearly seen in the figure~\ref{fig3P1} that the core-valence
correlation effects are saturated 
in the framework of the single and restricted double substitutions
and five layers of virtual orbitals, as
described in section~(\ref{Method}) above.
The remaining difference between the final Babushkin
  and Coulomb gauge values
may be attributed to the omitted core-core effects.
We have made an attempt to estimate the contribution of 
the core-core correlation to 
%
%
$ ^3 \! D _2 - \, ^1 \! S _0 $
and
$ ^1 \! P _1 - \, ^1 \! S _0 $
transition rates~\cite{BieronRa3d2004},
but for other transitions
the gauge differences and comparisons with data obtained by other
authors, where available, are
 the only indications of the accuracy
of our calculated rates.
Although the gauge difference must not be treated as the error bar per se,
it is a useful indicator in partially saturated multiconfiguration
 calculations of transition rates.
The values obtained in the Babushkin gauge are weighted toward the
outer parts of the electronic wavefunctions, while the Coulomb gauge values
weight more inner parts, where the core-core effects arise.
Therefore partially saturated expansions often produce 
Babushkin  and Coulomb gauge values converging toward different limits,
as in figure~\ref{fig3P1},
with the difference arising from the omitted core-core effects.
This is also the reason why Babushkin gauge results are usually
treated as more reliable, which seems to be confirmed by 
the agreement of the Babushkin gauge transition rate
 $A$($ ^3 \! P _1 - \, ^1 \! S _0 $)
with experiment
 (although agreement this close is very likely accidental;
  and good agreement for one level is not enough to justify
  a more general rule).

\subsection{The
 $ 7s6d ~ ^3 \! D _2 $
and
 $ 7s7p ~ ^1 \! P _1 $
states}
\label{3D2:1P1}

These two levels are distinguished because they were the subject
of a separate paper~\cite{BieronRa3d2004}.
In the present paper we duly quote the data obtained in core-valence
approximation,
and the reader is referred
to the above mentioned paper for further discussion.

\subsection{The remaining
$7s7p ~ ^3 \! P _2 $,
$7s6d ~ ^3 \! D _1 $, 
$7s6d ~ ^3 \! D _3 $, and
$7s6d ~ ^1 \! D _2 $
 states}
\label{3P2:1-3D1-3}

The remaining
$7s7p ~ ^3 \! P _2 $,
$7s6d ~ ^3 \! D _1 $, 
$7s6d ~ ^3 \! D _3 $, and
$7s6d ~ ^1 \! D _2 $
 states, together with the states discussed in
 sections~\ref{3P0},
~\ref{3P1}, and
~\ref{3D2:1P1} above,
constitute full set of states arising from the three lowest
electronic configurations of radium.

%
Some of these levels should also be considered metastable, not
only $ 7s7p~ ^3 \! P _0 $.
The $ ^3 \! D _3 - \, ^3 \! D _2 $  transition
is the strongest 'direct' decay channel
for the $ ^3 \! D _3 $ state,
but it is in fact very weak
(comparable to that of the
$ ^3 \! P _0 - \, ^1 \! S _0 $ transition).
The rates of other possible 'direct' decay channels
would be still smaller
(e.g.~M2 transition
$ ^3 \! D _3 - \, ^3 \! P _1 $ rate
 is comparable to that of the
$ ^3 \! D _2 - \, ^3 \! P _0 $  transition),
therefore the multiphoton or hyperfine-induced
transitions may also play a role in the
$ ^3 \! D _3 $ lifetime.
Similar considerations may in principle apply to the
$ ^3 \! P _2$,
$ ^1 \! D _2$, and
$ ^3 \! D _2$ states
(we did not pursue this issue, though).
 
%
%
%
%
%
%
%
\begin{table}
\caption{Calculated transition rates between nine lowest
 levels of radium [$s^{-1}$].
 Transition multipolarities are denoted by E1, E2, M1, M2; 
the Babushkin and Coulomb gauge values by (B) and (C), respectively.
HFS means hyperfine-induced.
Numbers in brackets represent powers of 10.
} 
\label{tableABC}  
\begin{tabular}{cllll}
\hline\noalign{\smallskip}
\multicolumn{2}{c}{transition} & This work & Ref.~\cite{DzubaGinges2006} 
                                             & Expt.~\cite{Scielzo2006}  \\
\hline\noalign{\smallskip}
%
$ ^3 \! P _0 - \, ^1 \! S _0 $ &  HFS+E1M1 & 2.935[-2] \\
%
$ ^3 \! P _1 - \, ^1 \! S _0 $ & E1(B) & 2.374[6] & 2.760[6] & 2.37(12)[6] \\
                     & \phantom{E1}(C) & 1.873[6] \\
%
$ ^3 \! P _1 - \, ^3 \! P _0 $ &    M1 & 1.334[-2] \\
%
$ ^3 \! P _1 - \, ^3 \! D _1 $ & E1(B) & 8.794[1] & 9.850[1] \\
                    &  \phantom{E1}(C) & 4.025[3] \\
%
$ ^3 \! P _1 - \, ^3 \! D _2 $ & E1(B) & 1.775[-3] & 1.572[-3] \\
                    &  \phantom{E1}(C) & 1.607[2] \\
%
$ ^3 \! P _2 - \, ^3 \! P _0 $ & E2(B) & 1.185[-2] \\
                    &  \phantom{E1}(C) & 1.243[-2] \\
%
$ ^3 \! P _2 - \, ^3 \! D _1 $ & E1(B) & 4.310[3] & 4.897[3] \\
                    &  \phantom{E1}(C) & 9.722[3] & \\
%
$ ^3 \! P _2 - \, ^3 \! D _2 $ & E1(B) & 4.602[4] & 5.204[4] \\
                    &  \phantom{E1}(C) & 1.109[5] \\
%
$ ^3 \! P _2 - \, ^3 \! D _3 $ & E1(B) & 1.044[5] & 1.234[5] \\
                    &  \phantom{E1}(C) & 4.201[5] \\
%
$ ^1 \! P _1 - \, ^1 \! S _0 $ & E1(B) & 1.793[8] & 1.805[8] \\
                    &  \phantom{E1}(C) & 1.795[8] \\
%
$ ^1 \! P _1 - \, ^3 \! D _1 $ & E1(B) & 3.282[4] & 4.195[4] \\
                    &  \phantom{E1}(C) & 5.222[4] \\
%
$ ^1 \! P _1 - \, ^3 \! D _2 $ & E1(B) & 9.793[4] & 2.646[4] \\
                    &  \phantom{E1}(C) & 1.441[5] \\
%
$ ^1 \! P _1 - \, ^1 \! D _2 $ & E1(B) & 3.241[5] & 3.194[5] \\
                    &  \phantom{E1}(C) & 5.875[5] \\
%
$ ^3 \! D _1 - \, ^3 \! P _0 $ & E1(B) & 1.390[3] & 1.529[3]  \\
                   &   \phantom{E1}(C) & 7.940[3] \\
%
$ ^3 \! D _2 - \, ^1 \! S _0 $ & E2(B) & 2.524[-1] & 3.032[-1] \\
                   &   \phantom{E1}(C) & 1.630[-1] \\
%
$ ^3 \! D _2 - \, ^3 \! P _0 $ &  M2   & 3.021[-13] \\
%
$ ^3 \! D _2 - \, ^3 \! D _1 $ &  M1   & 5.082[-4] \\
%
$ ^3 \! D _3 - \, ^3 \! D _2 $ &  M1   & 6.352[-3] \\
%
$ ^1 \! D _2 - \, ^1 \! S _0 $ & E2(B) & 2.710[1] \\
                   &   \phantom{E1}(C) & 2.271[1] \\
%
$ ^1 \! D _2 - \, ^3 \! P _1 $ & E1(B) & 6.960[2] & 7.722[3] \\
                   &   \phantom{E1}(C) & 1.224[3] \\ 
%
$ ^1 \! D _2 - \, ^3 \! P _2 $ & E1(B) & 5.930[0] & 7.973[0] \\
                   &   \phantom{E1}(C) & 7.535[0] \\ 
%
\noalign{\smallskip}\hline
\end{tabular}
\end{table}
%
\begin{table}
\caption{Calculated lifetimes of eight lowest excited
 states of radium, 
 compared with 
 data from other authors.
} 
\label{table.tau}  
\begin{tabular}{cllllll}
\hline\noalign{\smallskip}
 state         & This work   & Ref.~\cite{DzubaGinges2006}
                                 & Ref.~\cite{Dzuba2000}
                                     & Ref.~\cite{HafnerSchwarz1978}
                                         & Ref.~\cite{Bruneau1984}
                                             & Expt.~\cite{Scielzo2006}  \\
\hline\noalign{\smallskip}
$ ^3 \! P _0$ & 34(15) $s$  $^a$   &  & &  &  \\
%
$ ^3 \! P _1$ & 421 $ns$     & 362 $ns$     & 505 $ns$   
                                            & 420 $ns$ 
                                            & 250 $ns$ 
                                                           & 422(20) $ns$ \\
%
$ ^3 \! P _2$ & 6.46 $\mu s$ & 5.55 $\mu s$ & 5.2 $\mu s$  & & \\
%
$ ^1 \! P _1$ & 5.56 $ns$    & 5.53 $ns$    & 5.5 $ns$     & & \\
%
$ ^3 \! D _1$ & 719 $\mu s$  & 654 $\mu s$  & 617 $\mu s$  & & \\
%
$ ^3 \! D _2$ & 3.95 $s$     & 3.3 $s$      & 15 $s$       & & \\
%
$ ^3 \! D _3$ & 157 $s$ $^b$ &              &              & & \\
%
$ ^1 \! D _2$ & 1.37 $ms$ & 0.129 $ms$ $^c$ & 38 $ms$     & & \\
%
\noalign{\smallskip}\hline
\multicolumn{7}{l}{ (a)
 based on E1M1 and hfs-induced decay channels in
 the $^{223}_{\phantom{1}88}$Ra isotope} \\
\multicolumn{7}{l}{ (b)
 based on $ ^3 \! D _3 - \, ^3 \! D _2 $ magnetic dipole transition only;
 other possible decay channels neglected } \\
%
\multicolumn{7}{l}{ (c)
 calculated from transition probabilities quoted after
 Table~VI in Ref.~\cite{DzubaGinges2006}} \\
\end{tabular}
%
%
\end{table}
%

\section{Summary and conclusions}
\label{Summary}

Table~\ref{tableABC} presents calculated transition rates,
%
and the lifetimes are presented in table~\ref{table.tau}.
Coulomb gauge values of electric multipole transitions
from the table~\ref{tableABC}
were not used to obtain the lifetimes.
With the exception of the $ ^3 \! P _0 $ level,
the transition rates are given with 4 significant digits
and the lifetimes with 3 significant digits,
but that not necessarily reflects their accuracy.
As discussed in section~(\ref{3P1}) above,
the accuracy of our results is difficult to estimate,
but it is probably much worse than 3 or 4 digits,
particularly for very weak transitions.

%
%
The results of the present calculations may be considered as
fully converged in core-valence approximation
(convergence has indeed been observed for all transitions,
 similarly to that shown in figure~\ref{fig3P1}),
 with core-core
effects omitted.

With the exception of the
$ ^1 \! P _1 - \, ^3 \! D _2 $ and
$ ^1 \! D _2 - \, ^3 \! P _1 $ transitions
(we cannot offer any plausible explanation for these discrepancies),
our Babushkin gauge values are in reasonably good agreement
with the results 
of Dzuba and Ginges~\cite{DzubaGinges2006}.
An interesting feature 
is the large discrepancy between the results obtained
in Babushkin and Coulomb gauges for transitions connecting
closely-lying triplet $P$ and triplet $D$ states.
The B/C ratios
 (i.e. the ratios of Babushkin versus Coulomb gauge results)
turned out to be much closer to unity for those transitions
when the experimental energies
in the transition operator were replaced by theoretical ones.
This observation, together with inconsistent results of the
calculations of energies of excited states, had led
us~\cite{BieronRa3d2004} to suggest 
an experimental verification of radium data in
 Moore's tables~\cite{Moore1971}.
At that time we were not aware of the papers by
Eliav~{\sl et al}~\cite{Eliav1996} and
Landau~{\sl et al}~\cite{Landau2000},
where the excitation energies of radium were calculated
in the coupled-cluster approximation.
More recently these calculations were independently confirmed
within the framework of the CI+MBPT theory~\cite{DzubaGinges2006}.
In both cases good agreement with experiment had been achieved.
There are two principal differences between the methods and
approximations used in the abovementioned three papers
with respect to the methods and
approximations used in the present paper.
The transition rates were the primary targets
 of the present calculations, not the transition energies.
 Therefore we optimised the electronic
wavefunctions separately for each of the nine states of interest.
The core-valence correlation effects were fully accounted for.
The core-core correlation effects, which are less important in
calculations of
transition rates and hyperfine structures, were treated
in a very crude approximation in the cases of
 $ ^1 \! S _0 $, $ ^3 \! D _2 $ and $ ^1 \! P _1 $ states,
and were neglected for other states.
The calculations of energy level differences require well balanced
 orbital sets
and extensive multiconfiguration expansions. The results are usually in
better agreement with experiment if a common set of orbitals is
used for both states.
If the orbital sets are separately optimised, the transition energy
is obtained as a pure difference of the total energies of the two
states of interest. They are both several orders of magnitude
(five orders in case of radium) larger
than the transition energy itself, therefore 
our calculated transition energy values
may be less accurate than the calculated
transition rates and hyperfine structures.
Any further refinement of the present calculations would require
computer resources, which are currently unavailable.

%
\section{Acknowledgments}
\label{Acknowledgments}

\noindent
%
This work was supported by
 the Polish Ministry of Science and Higher Education (MNiSW)
in the framework of the scientific grant No.~1~P03B~110~30
awarded for the years 2006-2009.

\noindent
Laboratoire Kastler Brossel is Unit{\'e} Mixte de Recherche du CNRS
n$^{\circ}$ C8552.

\noindent
P.J.~acknowledges the support from the Swedish Research Council (VR).

\bibliography{ra-tau}

\end{document}